\documentclass[%
 reprint,
 amsmath,amssymb,
 aps,
floatfix,
]{revtex4-2}

\usepackage{graphicx}
\usepackage{dcolumn}
\usepackage{bm}
\usepackage{xcolor}

\begin{document}

\preprint{APS/123-QED}

\title{Ripplocations in Layered Materials: Sublinear Scaling and Basal Climb}

\author{James G. McHugh}
 \email{j.g.mchugh@lboro.ac.uk}
\author{Pavlos Mouratidis}
\author{Kenny Jolley}

\affiliation{%
 Dept. of Chemistry, Loughborough University\\
 Epinal Way, Loughborough LE11 3TU, United Kingdom
}%

\date{\today}

\begin{abstract}
The \textit{ripplocation} is a crystallographic defect which is unique to layered materials, combining nanoscale delamination with the crystallographic slip of a basal dislocation. Here, we have studied basal dislocations and ripplocations, in single and multiple van der Waals layers,  using analytical and computational techniques. Expressions for the energetic and structural scaling factors of surface ripplocations are derived, which are in close correspondence to the physics of a classical carpet ruck. Our simulations demonstrate that the lowest-energy structure of dislocation pile-ups in layered materials is the ripplocation, while large dislocation pile-ups in bulk graphite demonstrate multilayer delamination, curvature and voids.
This can provide a concise explanation for the large volumetric expansion seen in irradiated graphite.
\end{abstract}

\maketitle


\textit{Introduction.} --- It has long been understood that the plastic deformation of solids proceeds primarily through the nucleation of linear defects, called dislocations, which locally accommodate crystallographic slip. The slipped and unslipped crystal regions on either side of a dislocation line are related through translation by a lattice vector, which is a topological property of that dislocation, called its \textit{Burgers vector}, $\mathbf{b}$. The dynamics of dislocations are dominated by Frank's rule, which dictates that the energy of a dislocation is quadratic in the Burgers vector, $E \propto \mathbf{b}^2$. This causes perfect dislocations to dissociate into partials, with Burgers vector of less than a full lattice translation \cite{Frank1949,Frank1950,Frank1951}.

Recently, distinct behaviour has been observed in layered materials. Surface ripples with well-defined crystallographic character comprising sharp, localised folds between regions which are slipped relative to one another by an in-plane  lattice vector have been observed at the (0001) surface of exfoliated MoS\textsubscript{2} samples and graphite nanoplatelets \cite{Kushima2015,Alaferdov2018}. These defects combine the properties of edge dislocations and delamination, and have been termed \textit{ripplocations}. Extensive spherical nanoindentation experiments have 
demonstrated that the deformation behaviour of a wide variety of layered materials 
proceeds not through the basal dislocation, but through a distinct and reversible mechanism, which has been attributed to the formation of ripplocations \cite{Barsoum2017}.
Reversible arrays of ripplocation boundaries which can evolve into kink bands under increasing damage are believed to constitute the initial states of material failure in layered materials.
Ripplocations are now believed to occur on a variety of length scales, accounting for material deformation in substances as diverse as playing cards and steel sheets to phyllosilicates in the lithosphere \cite{Aslin2019,Barsoum2020,Barsoum2019}. 
Additionally, surface wrinkles on homogeneous substrates necessarily have edge dislocation character below a critical length \cite{Zheng2020}, hence the study of ripplocations will also give further insight into the wrinkling behaviour of 2D materials, which can substantially modify the physical properties of adsorbed monolayers. \cite{No2020, Tripathi2021, Zhu2012, Deng2016}

\begin{figure*}[htbp]
\includegraphics[width=0.5\textwidth, height=\textheight,keepaspectratio, trim = 8.5cm 6cm 8.5cm 0cm]{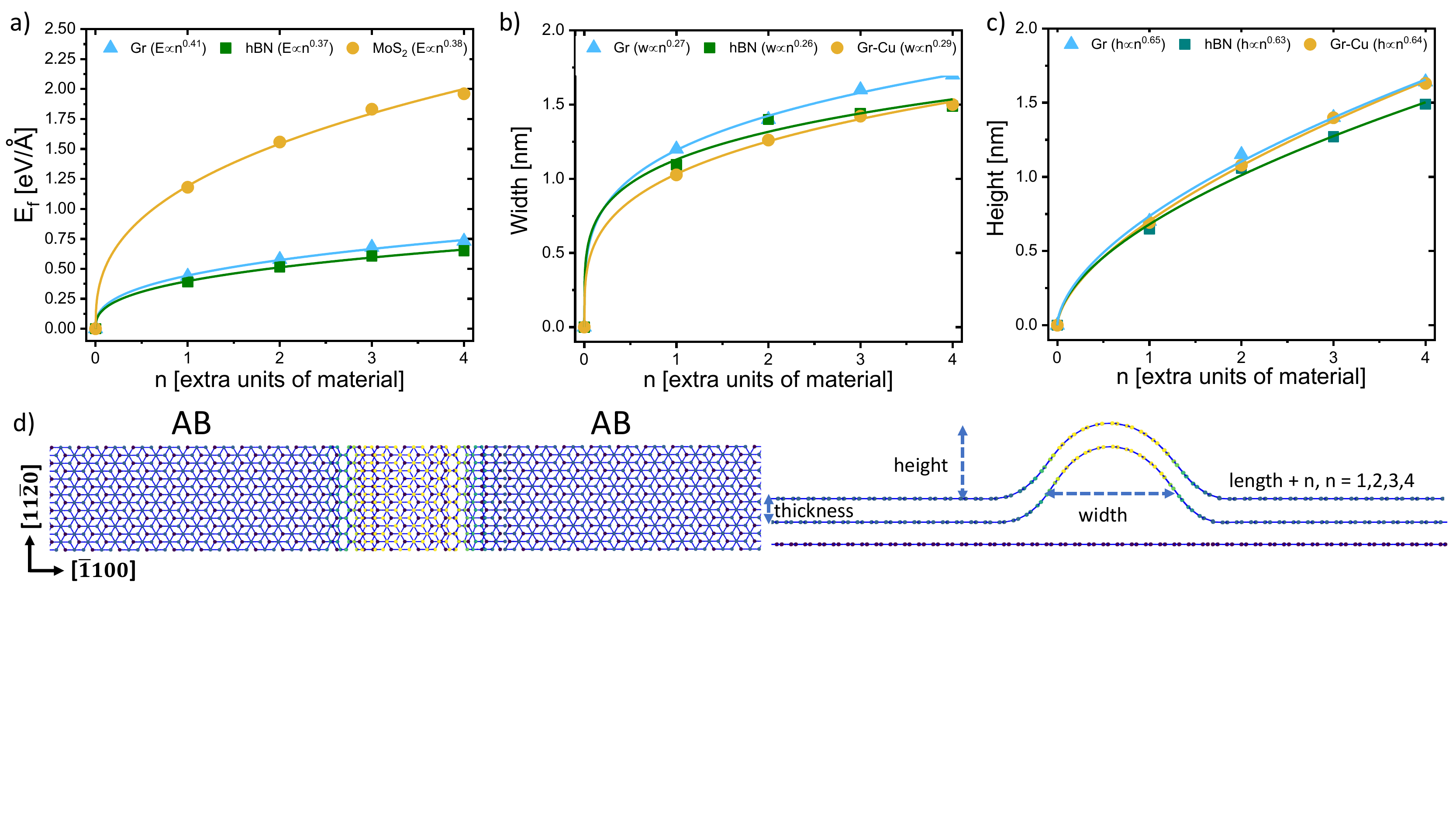}
\caption{(a) Formation energy of ripplocations in epitaxial graphene, MoS\textsubscript{2} and hBN sheets vs amount of additional material \textit{n}. (b) Width and (c) Height of ripplocations on epitaxial graphene on graphene, hBN on hBN and graphene on the Cu(100) surface (d) Schematic depiction of ripplocation configurations and associated structural properties as they have been calculated in this work. }
\label{fig1}
\end{figure*}

Ripplocations occur both epitaxially and in the bulk of layered materials. In the context of irradiated graphite, the bulk ripplocation or "ruck and tuck" mechanism of dislocation pile-up represents the first proposal of deformation due to delamination in the bulk of layered solids, and is likely to be particularly important in explaining many of the properties of dimensional change in highly-irradiated  highly oriented pyrolytic graphite (HOPG) samples. However, while this type of bulk ripplocation has been found in experimental TEM images of irradiated graphite \cite{Barsoum2020, Asthana2005, Eapen2013, Muto1999, Johns2020, Johns2020b, Pan2014}, it is a relatively rare defect. Distinct behaviour, including delamination and buckling spread across many layers and two-dimensional kink band networks \cite{Hinks2014, Hinks2014b}, has also been observed. This motivates further, systematic studies of multi-layer and two-dimensional dislocation configurations.

In contrast to the classical dislocation, ripplocations exhibit a \textit{sublinear} energy scaling as a function of Burgers vector, $E \propto \mathbf{b}^\beta$, 
where $\beta < 1$. This ensures that ripplocations comprised of multiple Burgers vector become increasingly energetically favourable as a function of extra material, as $|\textbf{b}_1 + \textbf{b}_2|^\beta < |\textbf{b}_1|^\beta + |\textbf{b}_2|^\beta$. While this sublinearity has been well-established, there is reasonable variation in the literature with reported values lying in the range $\beta \approx 0.3-0.45$ \cite{Gruber2016, Kushima2015, Ostadhossein2017}. Additionally, while there have been computational studies of ripplocations, no rigorous comparison has been made between them and the corresponding dislocation cores.


In this work, we have applied analytical and computational techniques which accurately model the interlayer friction between van der Waals layers to calculate the energetic and structural scaling factors of single- and multi-layer surface and bulk ripplocations. In van der Waals materials the interlayer friction is encapsulated in the $\gamma$-surface energy to slide adjacent layers along different crystallographic directions \cite{Dai2016,Kolmogorov2005,Popov2011,Wen2018}.
This is essential in accurately modelling these defects, as it is the resistance to sliding against adjacent layers that stabilises basal dislocations \& ripplocations. Our simulations of dislocation pile-ups in large systems demonstrate that the locally buckled ripplocation structure is the preferred mode of deformation with increasing additional material in the pile-up.


\textit{Methods.} ---  Density Functional Theory (DFT) calculations have been performed using plane-wave basis Kohn Sham states as implemented in the Quantum Espresso package \cite{Giannozzi2009, Giannozzi2017}. We have employed Vanderbilt ultrasoft pseudopotentials parameterised according to the Local Density Approximation (LDA). While it is often assumed that the inclusion of gradient terms in the Generalised Gradient Approximation (GGA) makes it a better approximation than the simpler LDA, in some cases LDA performs significantly better than GGA. Notably, the LDA provides a more accurate approximation to the interlayer van der Waals interaction in comparison to the GGA \cite{Garrity2014}, when benchmarked against higher levels of theory such as Quantum Monte Carlo (QMC) calculation. \cite{Nekovee2003,Mostaani2015}
Electronic wave-functions were expanded in a plane-wave basis with cut-offs of $E_{cut} = 40 \; Ry$.
A $\Gamma$-centred $1 \times 5 \times 1$ \textit{k}-point grid is found to sample the Brillouin zone with sufficient accuracy, giving total energy convergence to within $0.1$ meV/atom for all structures. Molecular dynamics (MD) simulations have been performed in LAMMPS \cite{Plimpton1995}, using  the hybrid neural network (hNN) potential for multilayer graphene systems developed by Wen and Tadmor \cite{Tadmor2011}. We have taken care to assess the suitability of the hNN potential and a variety of interlayer potentials in capturing the behaviour of dislocations and ripplocations (for additional details see Fig. S1-S2). 

\textit{Surface ripplocations.} --- Surface ripplocations have been created through the insertion of additional rows of material to the top monolayer of a bilayer structure in a periodic ribbon geometry, which is extended along one crystallographic orientation. Atoms in the top layer are allowed to fully relax, while the bottom layer is held fixed by setting all forces to zero, to approximate a compressed monolayer on a substrate. This approach has been verified in comparison to a fully-relaxed, multilayer graphite substrate using MD calculations, which produces very similar qualitative behaviour, with a small quantitative difference which we attribute to relaxation of the multilayer substrate (for more details see Fig. S3-S4). 

The amount of additional material is quantified using the parameter \textit{n}, denoting a Burgers vector of multiple lattice translations. Along the zigzag direction this corresponds to an edge dislocation ($n=1$) or superdislocation ($n>1$), 
$\textbf{b}_{\textbf{zz}} = n \times [1\overline{2}10]$.
Along the armchair direction $\mathbf{b}_{\textbf{ac}} = n \times [\overline{1}100]$, and all edge dislocations are superdislocations. Thus, \textit{n} quantifies the number of edge dislocations in the superdislocation pile-up, and is the characteristic property of a ripplocation defect. 

\begin{table*}
\small
  \begin{tabular*}{0.95\textwidth}{@{\extracolsep{\fill}}lllll}
    \hline
    \textbf{Method} & $\alpha_{best fit}$ & $E\propto b^{\beta(\alpha)}$  & $h\propto b^{\delta(\alpha)}$ & $w\propto b^{\gamma(\alpha)}$  \\
    \hline
Gravity   & 1.0       & $5/7$   & $4/7$ & $1/7$ \\
Gr (DFT)   & 0.191       & 0.401 {[}0.415{]}   & 0.588 {[}0.646{]} & 0.210 {[}0.292{]} \\
HBN (DFT)  & 0.139       & 0.388 {[}0.393{]}   & 0.629 {[}0.651{]} & 0.197 {[}0.303{]} \\
MoS\textsubscript{2} (DFT) & 0.127       & 0.378 {[}0.388{]}   & 0.610 {[}0.652{]}   & 0.287 {[}0.305{]}  \\
KC (MD)    & 0.173       & 0.402 {[}0.408{]}   & 0.623 {[}0.647{]}  & 0.187 {[}0.295{]}    \\
DRIP (MD)  & 0.197       & 0.411 {[}0.418{]}    & 0.617 {[}0.645{]}     & 0.226 {[}0.291{]}  \\
LP (MD)    & 0.170       & 0.382 {[}0.406{]}   & 0.549 {[}0.648{]}  & 0.109 {[}0.296{]}  \\
hNN (MD)   & 0.098       & 0.382 {[}0.413{]} & 0.679 {[}0.646{]} & 0.223 {[}0.293{]} \\
hNN t = 2  & 0.111       & 0.408 {[}0.381{]} & 0.759 {[}0.654{]} & 0.271 {[}0.309{]} \\
hNN t = 3  & 0.133       & 0.391 {[}0.391{]}   & 0.736 {[}0.674{]} & 0.287 {[}0.304{]}    \\
hNN t = 4  & 0.127       & 0.388 {[}0.388{]} & 0.765 {[}0.674{]} & 0.305 {[}0.306{]} \\
    \hline
  \end{tabular*}
    \caption{Energy, width, and height scalings of surface ripplocations from different computational methods. Corresponding values for the best-fit value of the sublinearity factor $\alpha$ are shown in square brackets.}
\end{table*}

DFT structural relaxations have been performed for zigzag surface ripplocations, $\mathbf{b} = n \times[1\overline{2}10]$ in graphene, hexagonal boron nitride (hBN) and MoS\textsubscript{2}. 
While the initial configuration is that of a superdislocation, with a well-defined Burgers vector, upon relaxation the large initial in-plane strains are released by out of plane buckling. This gives close to equilibrium bond lengths across the ripplocation (see Fig. S5-S6), and prior work has suggested that ripplocations do not possess a long range Burgers vector \cite{Kushima2015,Gruber2016}. 

Formation energies $E_f$ have been calculated from the optimised supercells as:
\begin{equation}
    E_f = E_{cell} - N \epsilon_{BL},
\end{equation}
where $E_{cell}$ is the energy of a defected bilayer, $\epsilon_{BL}$ is the energy per atom of a perfect bilayer, and N is the number of atoms. The height has been calculated as the distance between the flat part of the top layer and the ripple's peak, while the width was taken as the FWHM, both of which are shown schematically in Fig. 1(d).
The formation energy as a function of \textit{n} is shown in Fig. 1(a) for all three materials, where we observe the expected sublinear dependence on $\mathbf{b}$, with similar scaling factors in the range $\beta \approx 0.37-0.41$.

\begin{figure}[htbp]
\includegraphics[width=0.5\textwidth, height=\textheight, keepaspectratio, trim = 8cm 3.5cm 9cm 4.5cm, clip]{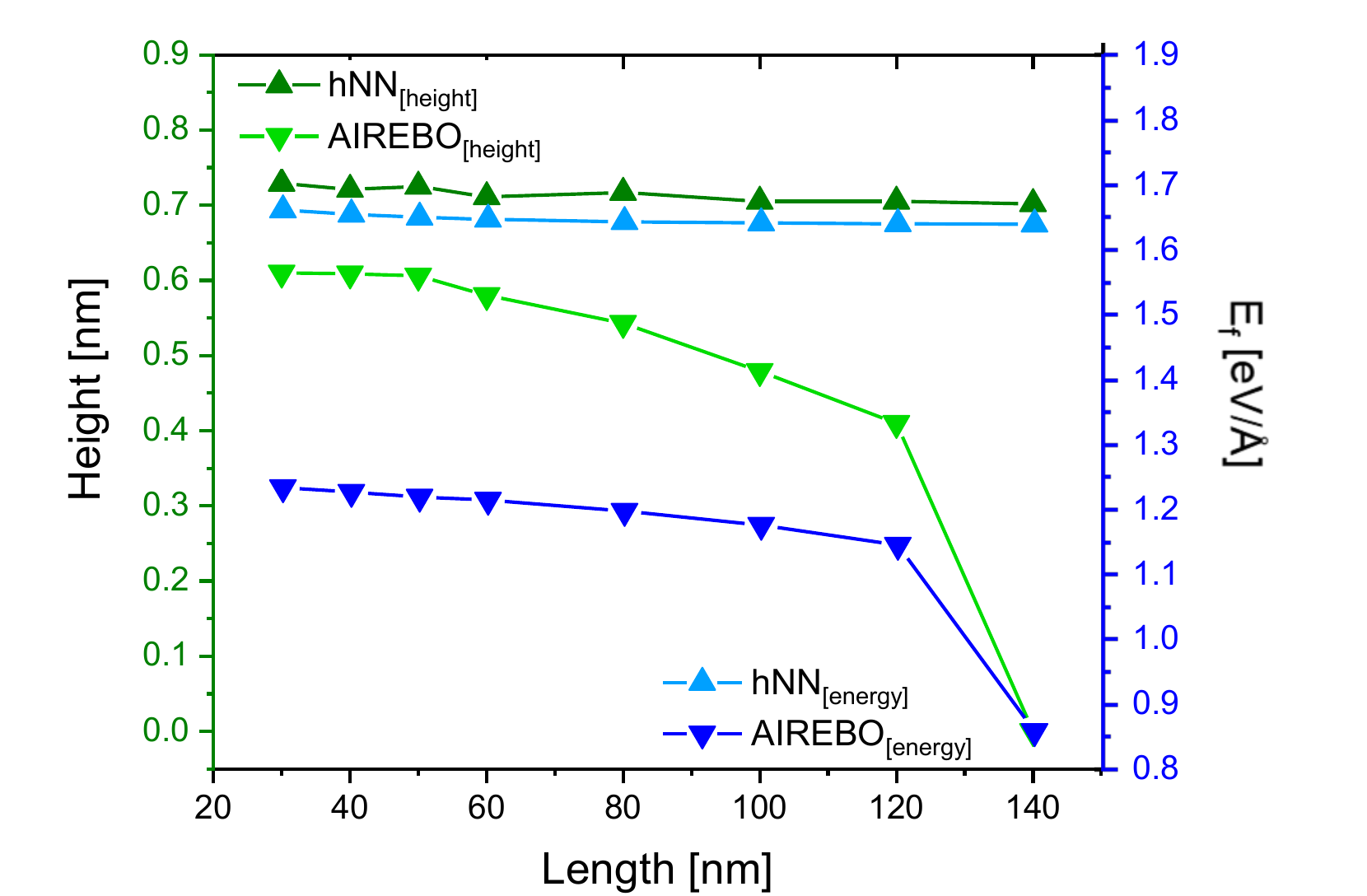}
\caption{Formation energy and height of the $n = 1$ surface ripplocation using the AIREBO and hNN potentials.}
\label{newfig1}
\end{figure}

Similarly, sublinear scaling is observed in the height and width, with consistent values for ripplocations in all three materials, as shown in Fig. 1(b) and Fig. 1(c). Further simulations of a graphene wrinkle on the Cu(100) surface (for additional details see Fig. S7-S8) produces nearly identical structural scaling factors, which is highly suggestive that similar physics underlies both types of ripples.

\begin{figure*}[htbp]
\includegraphics[width=1\textwidth, height=\textheight,keepaspectratio, trim = 1cm 0.5cm 1cm 0.5cm, clip]{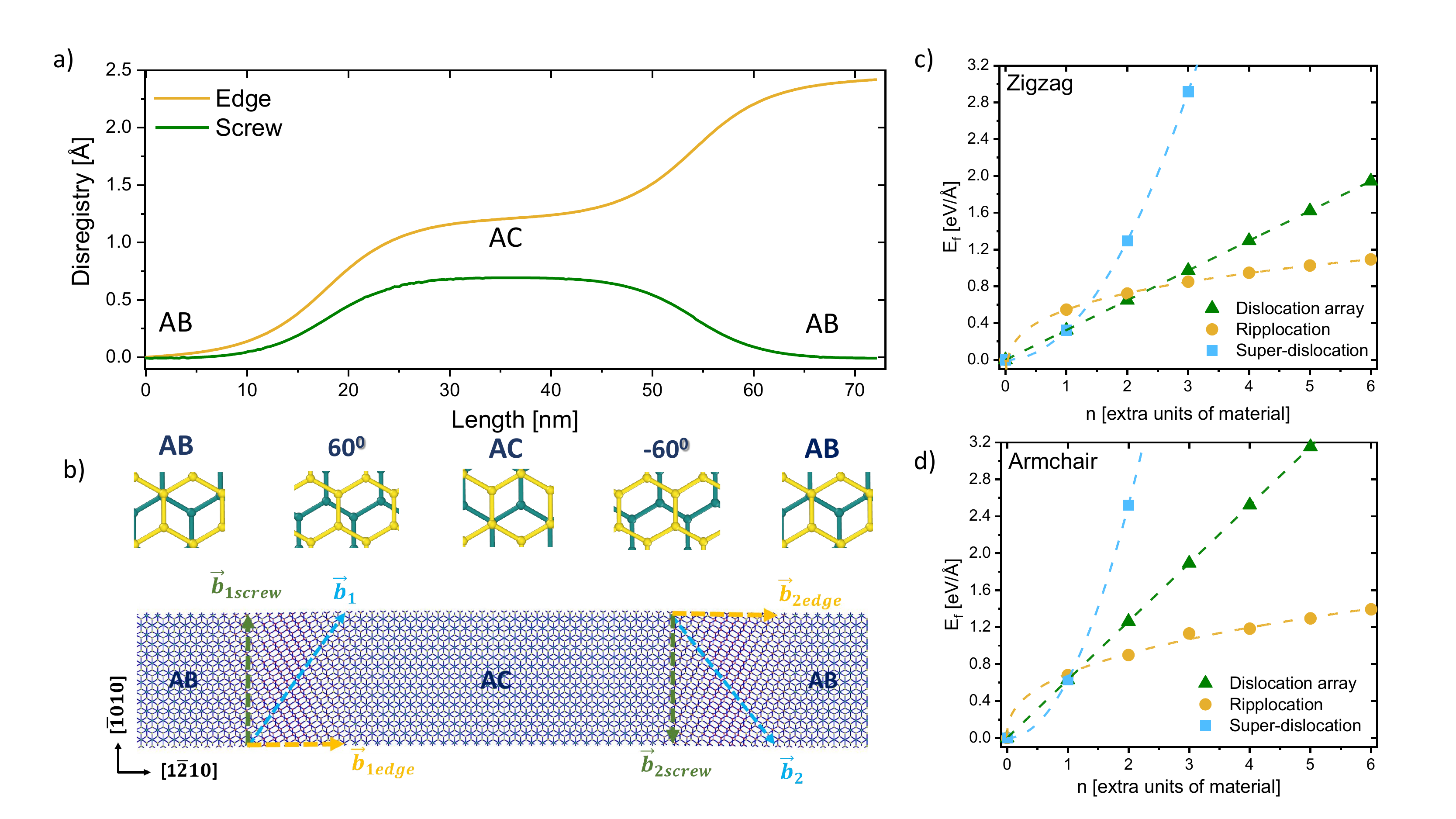}
\caption{(a) Disregistry calculated normal to the dislocation line (Edge) component and along the line (Screw) vs length of bilayer and (b) cell configuration and local stacking in a dislocated bilayer with two oppositely signed partial mixed dislocations (c) Comparison of superdislocation, dislocation array and ripplocation energies for Burgers vectors along to the zigzag and (d) armchair directions.}
\label{fig2}
\end{figure*}



\textit{Basal Dislocations} --- The sublinear energy scaling guarantees that ripplocations will eventually become lower in energy than the corresponding dislocation structures, and that with increasing Burgers vector, the dislocation core will eventually buckle and release its strain in the out of plane direction. However, the precise transition point between these two defects is unclear. In order to gain more detailed insight into the energetics and dynamics of ripplocations, we now compare the corresponding properties of basal dislocations. The basal dislocation core in graphite and other layered materials is wide ($>$10 nm) \cite{Alden2013, Lin2013, Butz2013} due to the weak interlayer interactions, and therefore difficult to simulate via DFT \cite{Lebedeva2016, Telling2003}.

To circumvent this, we have conducted MD simulations of large   cells of up to 500 nm in length. In conducting MD simulations of interlayer defects in layered materials, it is of is crucial importance to capture the interlayer friction or $\gamma$-surface energy. We have therefore taken particular care in comparing a variety of the available MD potentials to our DFT calculations. All of the potentials we consider agree qualitatively with our DFT results for the $\gamma$-surface energy \cite{Lebedeva2016, Kolmogorov2005, Wen2018, Wen2019}, as they are all fitted to quantum mechanical calculations of this property.

To further validate these potentials and compare to our DFT results, we have performed relaxation of surface ripplocations for increasing \textit{n}, along both armchair and zigzag directions. Our results for the energetic and structural scaling factors using the hNN potential are consistent with DFT, as are all of the other potentials which produce a significant interlayer friction energy. (for more details see Fig S9-S11). A full comparison of the scaling factors for all of the interlayer potentials considered, as well as the DFT results is shown in Table 1. We note that all of the computational methods predict highly sublinear energy scaling in agreement with DFT, and are generally within a relatively small range of values despite differences in choice of material, level of theory and computational method. Furthermore, simulation of multilayer (i.e. finite thickness, as shown schematically in Fig. 1 (d)) ripplocations, undertaken using the hNN potential also predict similar sublinear scaling factors.
 
In contrast, potentials which do not correctly capture the $\gamma$-surface energy, are not able to correctly capture ripplocation properties with increasing cell size. For example, while the AIREBO potential is commonly used in studies of graphene and graphite,  the $\gamma$-surface energy is significantly underestimated in comparison to DFT. This has immediate implications for the epitaxial properties of surface ripplocations. In particular, we find that this potential does not give consistent results (see Fig. 2). With increasing cell length, the ripple height decreases significantly, and for cells above approximately 55 nm ripplocations are not formed. Instead, the ripplocation ruck slides into the adjacent layers in a completely incommensurate manner (i.e., without forming basal dislocations).


Having verified the validity of our computational methods, we now compare the properties of ripplocations to the corresponding dislocation and super-dislocation cores. The structures considered have an identical Burgers vector, i.e. amount of extra material, in the core region but are constrained to remain flat and the two defects are therefore directly comparable. Fig. 3(c) and Fig. 3(d) compare the energy of ripplocations vs the corresponding superdislocation cores as a function of \textit{n}, calculated using the hNN potential.  It can be seen that at small \textit{n} for Burgers vector along both crystallographic directions, the lowest-energy structures are not localised ripples, but rather arrays of partial dislocations with net Burgers vector equal to the initial superdislocation Burgers vector, as shown in Fig. 3(a) and Fig. 3(b). 

For the zigzag core this gives pairs of $\pm 60^0$ partials, $\textbf{b}_{\textbf{zz}} = n\times1/3[1\overline{2}10] = n\times(1/3[1\overline{1}00] + 1/3[0\overline{1}10])$ for $n < 3$, above which ripplocations are nucleated in preference to dislocation arrays. Along the armchair direction, the initial $n=1$ superdislocation dissociates into a sequence of four partial dislocations, a pair of $\pm 30^0$ partials and a pair of edge partials, $\textbf{b}_{\textbf{ac}} =n\times\sqrt{3}[10\overline{1}0] = n\times(\sqrt{3}/6[1\overline{1}00] + \sqrt{3}/6[01\overline{1}0] + \sqrt{3}/3 [10\overline{1}0] + \sqrt{3}/3 [10\overline{1}0])$, and ripplocations are preferred for $n>1$.
Overall, at relatively small Burgers vector we observe the onset of rippling and buckling in preference to in-plane strain accommodation for both crystallographic directions.
This behaviour is again consistent with that of a ruck in a classical inextensible material, where under an initial compression, a sheet will not buckle unless the ratio of the initial compression to the coefficient of static friction of the substrate interaction, here encapsulated in the $\gamma$-surface, is above a critical value \cite{Vella2009}.

\textit{Scaling relations.} --- The near universality of the observed scaling law warrants further consideration. In prior work, the energetic and structural scaling factors for a ruck under the influence of gravity have been calculated. \cite{Vella2009, Kolinski2009}.
Taking into account our prior observation that the bond strain across a surface ruck is generally very low (Fig S5-S6), it is valid to consider a surface ripplocation as just such a ruck in an inextensible material. It is therefore noteworthy that our computational values are quite different from these classical scalings (see Table 1), particularly for the height and energy. 

We therefore revisit and modify these scalings, by considering two energetic contributions to the core energy \cite{Kushima2015, Vella2009, Kolinski2009}. We emphasise that inextensibility is an important condition in the following derivation, as it implies that the membrane force \cite{Lu2009} across the ruck is small, which allows us to ignore the contribution of strain to the ripple formation energy. This has been explicitly verified by calculating the strain contribution for the hNN model, which we find to be approximately $0.1$ eV and crucially is nearly constant as a function of \textit{n}. The total energy of a ripplocation ruck is then taken to be 
\begin{equation}
    E(w) = U_e(w) + U_s(w) = B \kappa^2 w + Vth^{\alpha}w.
    \label{energy_eq}
\end{equation}

The relevant energy terms comprise the elastic energy $U_e$ and the adhesion energy $U_s$, which is the energy acquired from the shifting and out of plane deformation across a ripple. The elastic energy is taken simply as $U_e = B \kappa^2w$ where B is the \textit{elastic bending energy}, $w$ is the width and $\kappa$ is the net curvature across a ripple. The interfacial term encapsulates the adhesion energy across the core, which we take to be $V th^\alpha w$ where $V$ is the interlayer adhesion energy of the absorbed layer to the substrate, $h$ is the height of the ruck and $t$ is the monolayer thickness.

The most important modification of this expression with respect to the classical equation for an inextensible ruck is the incorporation of the sublinearity factor $\alpha$, which accounts for the deviation from classical behaviour for van der Waals layers. We note that when $\alpha = 1$ this corresponds to the classical rucking of a sheet under the influence of gravity. 
Making use of the relations $h = \sqrt{b w}$ and $\kappa = \frac{h}{w^2}$ for the arc-length and curvature  of a clamped elastica, the energy can be expressed as a function of width only, $E(w) = Bbw^{-2} + Vtb^{\alpha/2} w^{1 + \alpha/2}$.
We then solve for the equilibrium width by considering the derivative of the energy equation, Eq. (2) with respect to the defect width,
\begin{equation}
    \frac{dE(w)}{dw} = -\frac{2 B b}{w^3} + (1+\alpha/2) Vt (b w)^{\alpha/2}  = 0.
\end{equation}
This immediately leads to analytical forms for the scaling of ruck height and width as $w(b) \propto b^{\gamma(\alpha)}$ and $h(b) \propto b^{\delta(\alpha)}$, with 
\begin{equation}
    \gamma(\alpha) = \frac{2 - \alpha}{6 + \alpha},
\end{equation}
and
\begin{equation}
    \delta(\alpha) = \frac{1}{2} + \frac{2 - \alpha}{12 + 2\alpha}.
\end{equation}

Replacing the height and width for these expressions in  Eq. 2, the ruck energy as a function of width is then $E(w) = (B + tV) A(\alpha)^\frac{2+\alpha}{6+\alpha} b^{1 - (4 - 2 \alpha)/(6 + \alpha)}$, where $A(\alpha) = \frac{2B}{tV(1+\alpha/2)}$, hence leading to a sublinear energy scaling 
\begin{equation}
    \beta(\alpha) = 1 - \frac{4 - 2 \alpha}{6 + \alpha}.
\end{equation}
The sublinear scaling as a function of height can be understood as a consequence of the shorter range of the  van der Waals interaction, which scales as a function of distance as $r^{-6}$, here incorporated approximately through the sublinearity factor $\alpha$. This leads to the fact that material closer to the substrate matters proportionally much more than that which is further away, 
so that the overall energy scales very slowly in the maximum height. 

Notably, our calculations indicate very similar scaling factors across all materials and approximations. For example, our calculations indicate that the interlayer adhesion energy factor $V$ of MoS\textsubscript{2} is around $50\%$ higher than that of graphene, yet the calculated scalings remain in a very similar range, indicating that it is indeed the short-ranged behaviour of the interlayer interaction which determines ruck properties, rather than the magnitude of the adhesion energy.

Thus, if the introduction of a sublinear term is an accurate model of the rippling across a ruck in a two-dimensional monolayer, all three properties will be consistently inter-related through a common sublinearity factor. 
Table 1 shows the calculated scaling of energy, height and width for ripplocations in different materials and using different methods, as well as a "best fit" value of $\alpha$. This value minimises the difference, $\epsilon$, between the analytical and computational values for all three scalings, $\epsilon = \sqrt{ (\beta_{fit} - \beta_{comp})^2 + (\delta_{fit} - \delta_{comp})^2 + (\gamma_{fit} - \gamma_{comp})^2 }$.
The analytical expression produces extremely good agreement with the computational sublinear energy dependence, with small errors in the height and width in the range of $0.05-0.1$. We attribute this small departure from the ideal behaviour for the structural properties to the violation of the $h = \sqrt{bw}$ assumption for our van der Waals systems (see Fig. S12)

Overall, our calculations confirm that all three scalings are indeed closely related.Additionally, since we have only considered the bending and adhesion energies, this scaling behaviour is immediately applicable to physisorbed monolayers on other substrates, this explains why our DFT simulation of the graphene/Cu(100) surface interaction produces scalings which agree closely to the ripplocation results in Fig. 1(b) and 1(c).

\begin{figure*}[t!]
\includegraphics[width=1\textwidth, height=\textheight,keepaspectratio, trim = 0cm 5cm 0cm 0cm, clip]{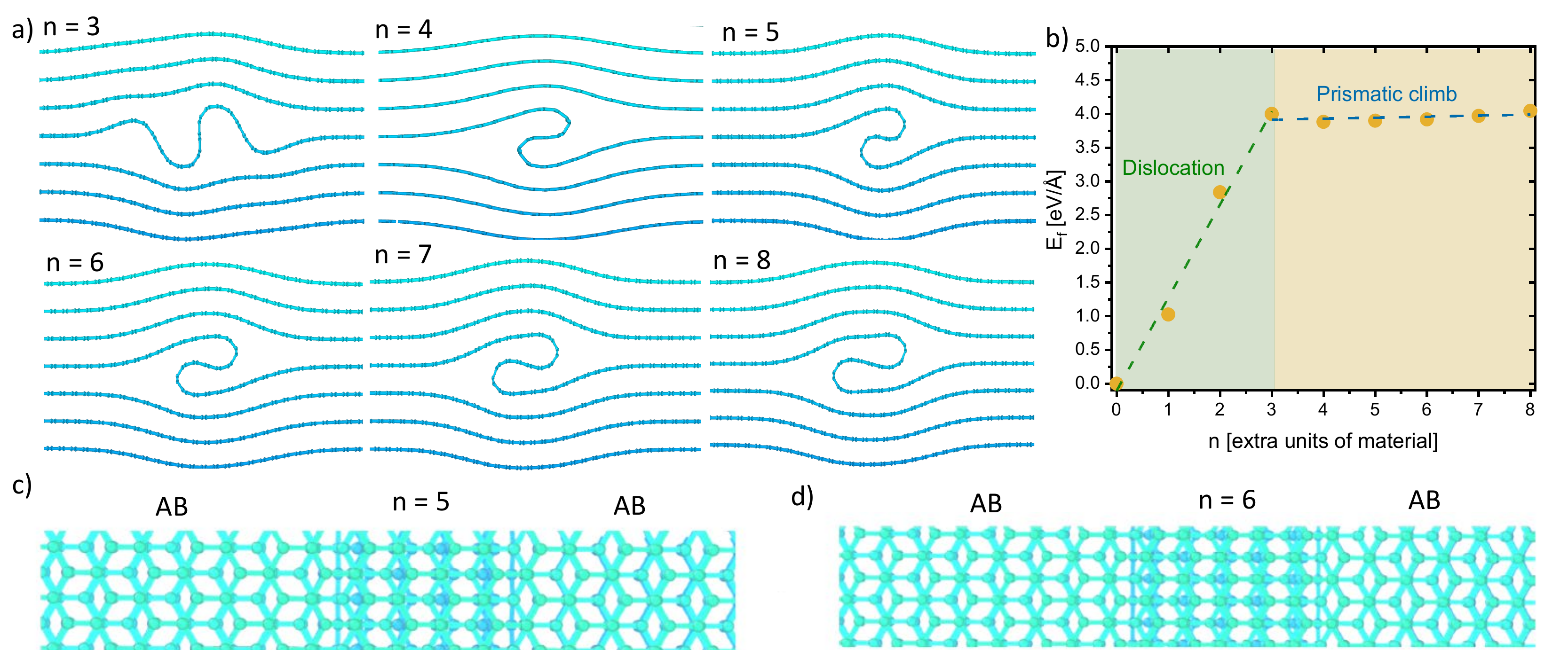}
\caption{(a) Single-layer pileup in bulk graphite for different values of \textit{n}, resulting in the "ruck and tuck" structure with increasing material, (b) bulk ripplocation formation energy vs \textit{n}, (c) and (d) plan view above a ruck and tuck defect, demonstrating that Bernal stacking is recovered in the material surrounding the folded defect core.}
\label{fig4}
\end{figure*}

\textit{Bulk.} --- We now discuss the behaviour of bulk ripplocations in large MD cells. Cells containing bulk ripplocations are created similarly to surface ripplocations, where extra rows of material are inserted into a monolayer in a bulk cell. For small \textit{n}, this again gives dislocation arrays and a linear energy dependence.  Above a critical value of \textit{n}, the formation energy is almost constant with additional material as shown in Fig. 4(a), and the folded-over bulk ripplocation, or "ruck \& tuck" structure is nucleated \cite{Heggie2011}. Accurate $\gamma$-surface energy is again paramount in attaining accurate defect structure, which for bulk ripplocations is reflected in the surrounding sheets, which have the lowest-energy Bernal stacking (see Fig. 4c-d), from which we conclude that slip of the folded defect into the surrounding layer is inhibited due to friction. 

The constant formation energy can be understood with reference to the relaxed structures, shown in Fig. 4(a). As extra material is added, it is accommodated by increasing the length of the prismatic sheet bounded by the highly curved regions, while the curved regions remain approximately the same. This demonstrates that for both bulk and surface ripples, curvature can essentially be seen as the mechanism of dislocation climb for basal dislocations in layered materials. This proceeds unconstrained on the surface, while in bulk additional material is accommodated simply by growing the flat, perfectly-stacked prismatic layer.

\begin{figure*}[t!]
\includegraphics[width=1\textwidth, height=\textheight,keepaspectratio, trim = 0cm 8cm 0cm 0cm, clip]{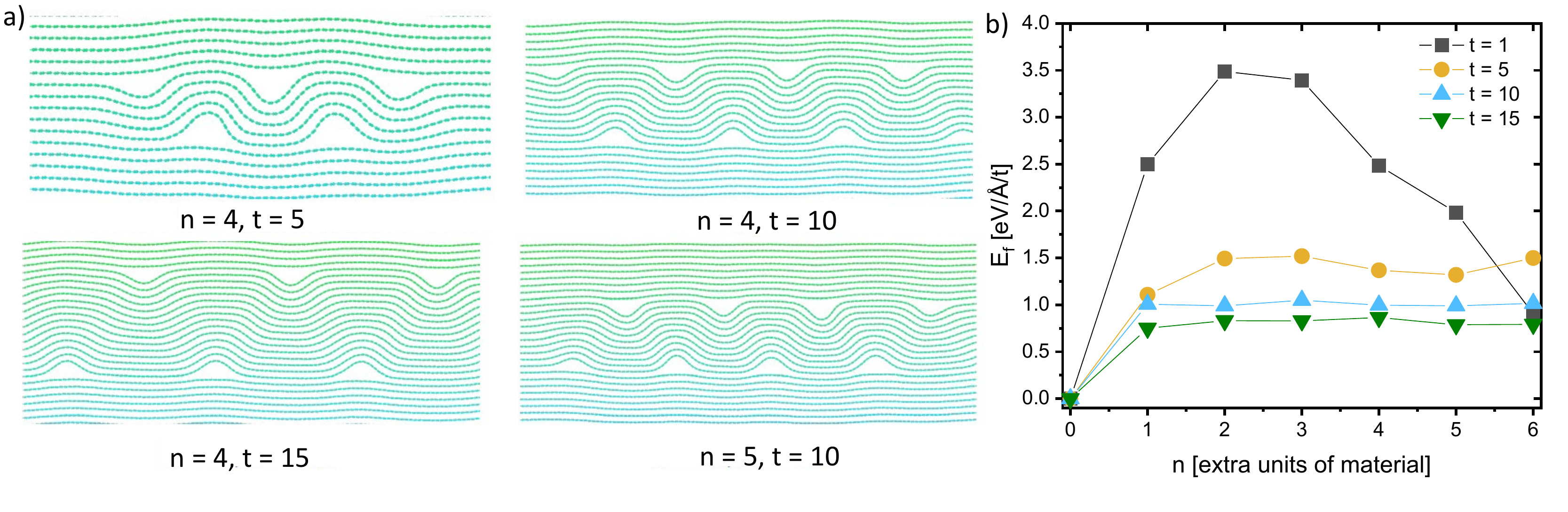}
\caption{(a) Multilayer (finite thickness, \textit{t}), dislocation pile-ups (b) Energy per length of additional material of multilayer dislocation pile-ups.}
\label{fig5}
\end{figure*}



Dislocation pile-ups on multiple adjacent layers in bulk graphite have also been considered through the incorporation of extra material to a number of adjacent layers in a large bulk simulation cell. These structures are the corresponding bulk defect to the finite-thickness surface ripplocations.
Upon relaxation we find structures which are quite distinct from the single-layer ruck and tuck, which are more similar to bulk twin boundaries \cite{Rooney2018}, demonstrating multilayer curvature and the formation of voids adjacent to the compressed region, some representative structures of which are shown in Fig. 5(a). The distinction between the single and multilayer defects can be understood through consideration of the bending energy for multilayer graphene, which scales linearly for small thicknesses \cite{Han2019}

The formation energy per Burgers vector per layer for these defects is shown in Fig. 6(b), which we see results in even
lower energy per \textit{n} than the single-layer ruck and tuck. The ruck and tuck can then be seen as a limiting case where compression occurs on only one layer. It is therefore unsurprising that the structure of irradiated HOPG generally involves larger-scale curvature, as the higher bending energy inhibits the ruck and tuck for the lower-energy, finite thickness bulk ripplocations. However, the presence of both the ruck and tuck defect as well as large-scale bending can be taken as evidence for dislocation pile-up as the deformation mode of irradiated HOPG.  Furthermore, all of the bulk structures simulated in this work essentially demonstrate basal climb, i.e., c-axis expansion to release in-plane compression resulting from many basal dislocations.

\textit{Two-dimensional ripples.} --- MD simulations also permit calculations of other large structures which are prohibitive for DFT, such as the two-dimensional ripplocation junction shown in Fig. 6. This has been created by inserting material along both the armchair and zigzag directions on the surface of a bilayer cell. 
The pictured cell, which has an initial Burgers vector $\mathbf{b} = (b_{ac}, b_{zz}) = (2,1)$, results in the formation of a ripplocation junction, where perpendicular dislocation lines meet at an AA-stacked dislocation node. It is noteworthy that this type of structure is strikingly similar to experimental transmission electron microscopy (TEM) images of irradiated HOPG surfaces \cite{Hinks2014, Hinks2014b}, and in this context our large-scale simulations again provide evidence of dislocation pile-up as the likely deformation mode of irradiated HOPG. We leave further, systematic investigation of these two-dimensional defects to future work.

\textit{Discussion.} --- The application of classical mechanics to the nanoscale buckling of two-dimensional monolayers has typically proven difficult. In this work, we have found that minor modification of classical expressions gives exceptional agreement to DFT calculations.
The energy scaling, Eq. \ref{energy_eq}, is based only on classical expressions for clamped elastica and is, therefore, directly applicable to any inextensible layered material, through simple modification of the $\alpha$ sublinearity parameter. This reinforces the idea that ripplocations are a general concept applicable at a variety of length-scales, since we have considered only monolayer bending and adhesion.

The exceptionally low sublinear factors, in the range $\alpha \approx 0.1 - 0.18$, in comparison to a sheet under the influence of gravity, helps to explain the exaggerated tendency of epitaxial monolayers to wrinkle and ripple.
Our work therefore indicates that the most important property in the wrinkling of epitaxially-grown monolayers is the coefficient of friction between the monolayer and the substrate. This can explain, for example, the high density of wrinkle formation in graphene adsorbed on the the highly incommensurate Cu(100) and Cu(110) surfaces \cite{Deng2017} in comparison to the almost-commensurate Cu(111) face. It is therefore noteworthy that the ripplocation formation energy is close to the observed experimental energy barrier for wrinkle nucleation on Ir(100) substrates. \cite{Hattab2012,NDiaye2009}

\begin{figure}[htbp]
\includegraphics[width=0.48\textwidth, height=\textheight, keepaspectratio, trim = 4.5cm 0cm 3.8cm 0cm, clip]{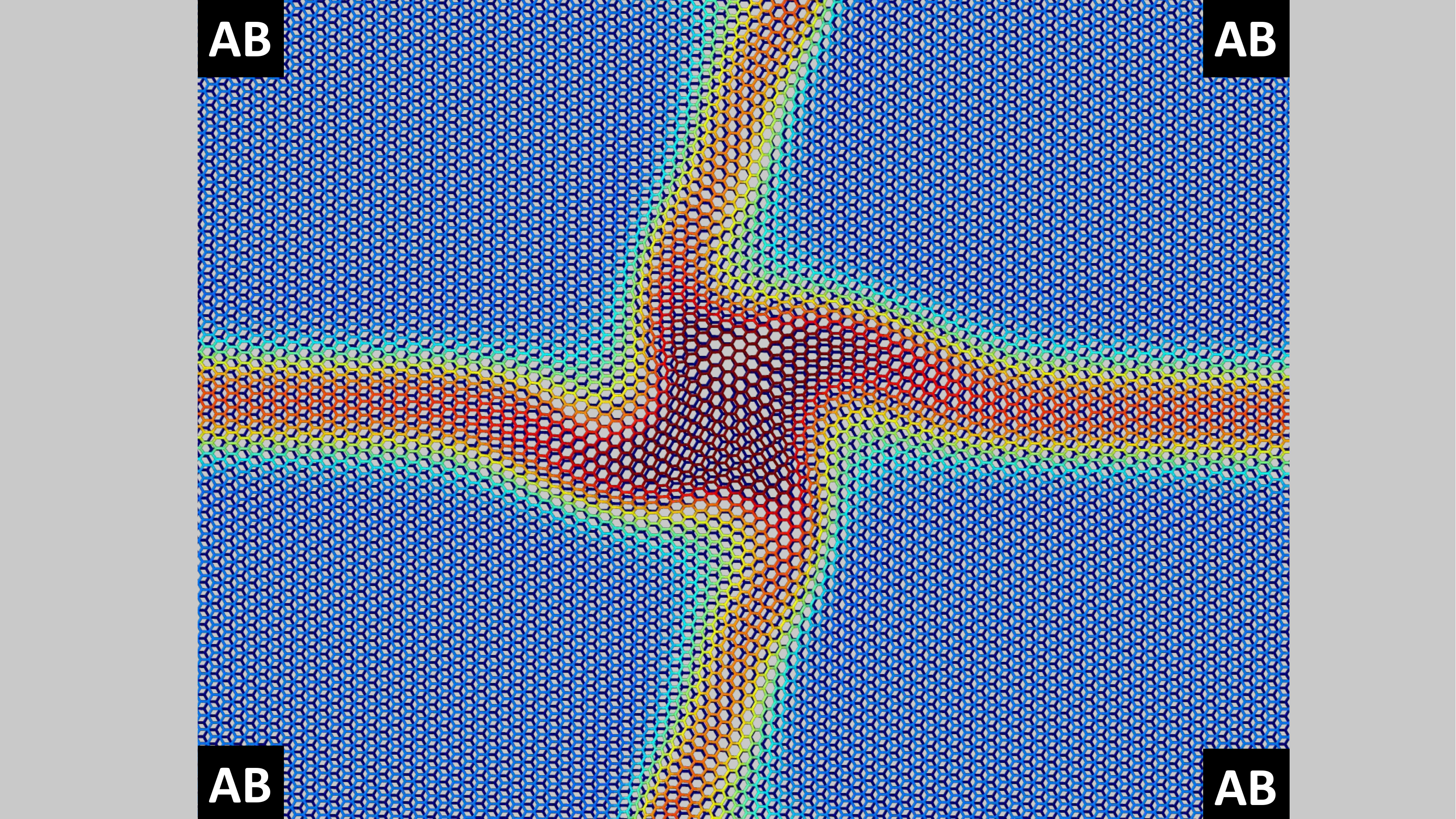}
\caption{wo-dimensional ripplocation junction on the Graphite (0001) surface. Carbon atoms are colour-coded according to height. Blue sections are flat AB stacked regions. The ripples height is around 1 nm, red sections.}
\label{fig6}
\end{figure}

In all cases, buckling of the ripplocation can be seen as a type of climb mechanism for basal dislocations. Our work therefore has important implications for the irradiation-induced deformation of HOPG. Under irradiation, HOPG and even single graphene sheets will shrink, which very quickly leads to the formation of basal dislocations.
The very low Peierls barrier in layered materials ensures that the basal dislocation core is very diffuse and easy to move, which readily allows dislocation pile-ups \cite{Telling2003,Heggie2011}.
Under this compression, our work confirms that these dislocation pile-ups will result in delaminated, curved regions of microstrucure rather than flat superdislocation cores. This suggests that essentially a process of sub-surface wrinkling is implicated in the pronounced c-axis expansion of irradiated HOPG samples, with increasing compression resulting in the formation of ripplocation arrays, voids, ruck and tucks and highly curved regions. 
It is then unsurprising that there are substantial similarities between our simulations and experiment. Under heavy ion bombardment, in-plane dislocations are nucleated, which evolve into rippled "kink boundaries". These two-dimensional dislocations complexes are almost identical to our simulations of surface ripplocation arrays, Fig. 6.

There is also substantial experimental evidence for the formation  of the single-layer ruck \& tuck in irradiated graphite, \cite{Barsoum2020, Asthana2005, Eapen2013, Muto1999, Johns2020, Johns2020b, Pan2014}. In addition, our multilayer simulations 
are similar to the regions of curvature and voids, which have recently been found in irradiated HOPG \cite{Johns2019}, again demonstrating damage due to compression and pile-up. In sum, this correspondence 
implicates the compression-induced dislocation pile-up and rippling of graphene monolayers as a concise and appealing explanation of the irradiation-induced deformation of single crystal graphite.

\textit{Acknowledgments} --- JGM was supported by the UK EPSRC grant EP/R005745/1. 
KJ and PM gratefully acknowledge funding from EDF Energy. 
The authors acknowledge the use of the HPCMidlands+ facility, funded by EPSRC grant EP/P020232/1.

\nocite{*}

\bibliographystyle{unsrt}
\bibliography{apssamp}
\end{document}